\documentclass[11pt]{article}
\usepackage{latexsym}  % for Symbol fonts
\usepackage{amssymb}
\usepackage{graphicx}
\usepackage{amsmath}
\usepackage{mathrsfs}
\newcommand{\be}{\begin{equation}}
\newcommand{\ee}{\end{equation}}
\newcommand{\bea}{\begin{eqnarray}}
\newcommand{\eea}{\end{eqnarray}}
\newcommand{\bref}[1]{(\ref{#1})}

\newcommand{\pa}{\partial}

\newcommand{\bomega}{\boldsymbol{\Omega}}
\begin{document}
%%%%%%%%%%%%%%%%%%%%%%%%%%%%
\begin{titlepage}
%%%%% PREPRINT NUMBERS %%%%%%
%\begin{flushright}
%\today
%\end{flushright}
%%%%%%%%%%%%%%%%%%%%%%%%%%%%%%
\vspace{4\baselineskip}
%%%%%%%%%%%%%%%%%%% TITLE %%%%%%%%%%%%%%%%%%
\begin{center}
{\Large\bf General Spin Precession and Betatron Oscillation\\ in Storage Ring}
\end{center}
%%%%%%%%%%%%%%%% AUTHORS %%%%%%%%%%%%%%%%%%%%%%%
\vspace{1cm}
\begin{center}
{\large Takeshi Fukuyama
\footnote{E-mail:fukuyama@se.ritsumei.ac.jp}}
%and
%{\large Alexander J. Silenko$^{b,c,}$
%\footnote{E-mail:alsilenko@mail.ru}}
\end{center}
%%%%%%%%%%%%%%%%%%%%%%% AFFILIATION %%%%%%%%%%%%
\vspace{0.2cm}
\begin{center}
{\small \it Research Center for Nuclear Physics (RCNP),
Osaka University, Ibaraki, Osaka, 567-0047, Japan}\\[.2cm]

%${}^{b} $ {\small \it Institute of Nuclear Problems, Belarusian State University, Minsk 220030, Belarus}

%${}^{c} $ {\small \it Bogoliubov Laboratory of Theoretical Physics, Joint Institute for Nuclear Research,
%Dubna 141980, Russia}

%\\and\\
%${}^{c}$ {\small \it Maskawa Institute for Science and Culture,
%Kyoto Sangyo University, Kyoto 603-8555, Japan}
%\medskip
\vskip 10mm
\end{center}
PACS number: 03.65Pm, 04.20.Jb, 11.10.Ef
\vskip 10mm

\begin{abstract}
We give the geralized expression of spin precession of extended bunch particles having both anomalous magnetic and electric dipole moments in storage ring.
The transversal betatron oscillation formula of the bunch is also given. The latter is the generalization of the Farley's pitch correction \cite{Farley}, including radial oscillation as well as vertical one. Some useful formulae for muon storage ring are discussed in appendix.
\end{abstract}
\end{titlepage}
%%%%%%%%%%%%%%%%%%%%%%%%%%%%%%%%%%%%%%%%%%%%%%%%
\section{Introduction}
The detailed analyses of spin precession of particles in storage ring are essentially important for the measurement of anomalous magnetic dipole moment (anomalous MDM or $g-2$) and electric dipole moment (EDM) of those particles.
The accelerated particles enter into the storage ring (in $x$-direction) in bunch , which have not only some transversal size ($y,z$-plane) but also energy width. These transversal and energetic extents in constant vertical ($z$-direction) magnetic field cause the transversal and longitudinal oscillations ($x$-direction) around the horizontal circular motion, respectively.
These oscillations are called betatron oscillations. In this letter, we discuss the transversal oscillation. Corresponding to this, spin precession formula is generalized to invoke the bunch profile. Pitch correction needs also generalization of the Farley's correction which discussed only vertical oscillation. 
This paper is organized as follows. In section 2, we discuss on the generalized Thomas-Bargman-Michel-Telegdi (BMT) equation. "generalized" has the duplicate meaning that particles have permanent EDM as well as anomalous MDM and that particles have the transversally extended profile, reflecting the realistic experiments. Transversal betatron oscillation is dicscussed in section 3.
\section{Generalized BMT equation of extended bunch particles}
Before the arguments of bunch particles, we make a brief comment on the BMT equation \cite{Thomas, BMT} of a particle.
In the previous paper we generalized the BMT equation in the presence of permanent EDM \cite{fuku1}. The resulting angular velocity of spin precession is
\bea
\bomega_s &=&-\frac{e}{m}\left[\left(a+\frac{1}{\gamma}\right){\bf B}-\frac{\gamma a}{\gamma+1}({\bf v}\cdot{\bf B}){\bf v}-\left(a+\frac{1}{\gamma+1}\right){\bf v}\times{\bf E}\right.\nonumber\\
&+&\left.\frac{\eta}{2}\left({\bf E}-\frac{\gamma}{\gamma+1}({\bf v}\cdot{\bf E}){\bf v}+{\bf v}\times {\bf B}\right)\right]
\label{Nelsonh}
\eea
in $\hbar=c=1$ units.
Here $a=\frac{g-2}{2}$ and $g$ and $\eta$ are defined as $g=2\mu m/(es),~\eta=2dm/(es)$. Here we consider the case $s=1/2$
and use the system of units $c=\hbar=1$ except for the concrete numerical estimation case.
This equation itself has been known before our work using the dual transformation 
\be
{\bf B}\rightarrow {\bf E},~{\bf E}\rightarrow -{\bf B},  ~\mbox{and}~a\rightarrow \eta/2
\ee
However, it is not so simple. Indeed, in its factor some ambiguity had appeared \cite{Semertzidis, Khriplovich} and there is the famous "Thomas half" problem even in the normal magnetic moment. This equation should have been derived explicitly as was done in \cite{fuku1}.

Since it is very useful for the systematic derivation of the final results of the present letter, we give more comments on a spin precession of a particle. 
One usually considers the spin motion relative to the beam direction. The beam motion obeys
\be
\frac{d{\bf v}}{dt}=\frac{e}{m\gamma}\left[{\bf E}+{\bf v}\times {\bf B}-{\bf v}({\bf v}\cdot{\bf E})\right],~~\frac{d\epsilon}{dt}=e{\bf v}\cdot {\bf E},
\label{eqm}
\ee
 and rewrites Eq.(\ref{eqm}) in terms of the unit vector in direction of the velocity (momentum), ${\bf N}={\bf v}/v={\bf p}/p$:
\be
\frac{d{\bf
N}}{dt}=\frac{\dot{\bf v}}{v}
-\frac{\bf v}{v^3}\left({\bf v}\cdot\dot{\bf v}\right)=\bomega_p\times{\bf N}, ~~~ \bomega_p=\frac{e}{m\gamma}\left(\frac{{\bf
N}\times{\bf E}}{v}-{\bf B}\right),
\ee
where $\bomega_p$ is the angular
velocity of rotation of the velocity, momentum, and beam directions. Thus, the angular velocity of the spin rotation relative to the beam direction (prticle rest frame) is given by
\be
\bomega'=\bomega_s-\bomega_p=-\frac{e}{m}\left[a{\bf B}-\frac{\gamma a}{\gamma+1}({\bf v}\cdot{\bf B}){\bf v}-\left(a-\frac{1}{\gamma^2-1}\right){\bf v}\times{\bf E}+\frac{\eta}{2}\left({\bf E}-\frac{\gamma}{\gamma+1}({\bf v}\cdot{\bf E}){\bf v}+{\bf v}\times {\bf B}\right)\right].
\label{Nelsonf}
\ee
This is suitable in the Frenet-Serret coordinate system. On the other hand, it may be more
useful to use the cylindrical coordinate in the case of the storage rings, where particles move in horizontal plane on average and the deviation from it is approximated small (Fig.1).
\begin{figure}[h]
\begin{center}
\includegraphics[scale=2.0]{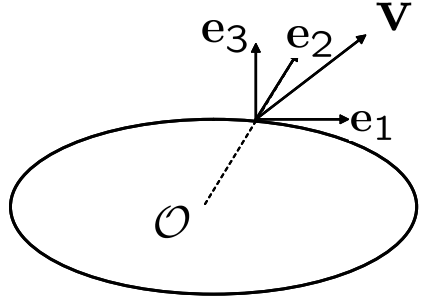}
\end{center}  
%\vspace*{-0.5cm}
\caption{\label{fig:geom_config}
The configuration of betatron oscillation. $x({\bf e}_1),~y({\bf e}_2)$ coordinates are those projected on the averaged plane (horizontal place) of the storage ring. z is vertical to the horizontal plane. ${\bf v}$ is the muon velocity (emphasized to see the small pitch correction) which has both y, z components with main x component.}
\end{figure}

In this frame, 
\be
\frac{d{\bf e}_1}{dx}=-\frac{{\bf e}_2}{\rho},~~\frac{d{\bf e}_2}{dx}=\frac{{\bf e}_1}{\rho},~~\frac{d{\bf e}_3}{dx}=0.
\ee
Here ${\bf e}_1$ is the tangential unit vector (x-component) to the beam's averaged circular motion in the horizontal plane (spanned by x,y coordinates), 
${\bf e}_2$ is the radial unit vector (y-component) in the horizontal plane and ${\bf e}_3$ is the vertical unit vector (z-component) to the horizontal plane.
$\rho$ is the radius of the averaged circle and
\be
\frac{1}{\rho}=\left|\frac{eB_z}{\gamma mv_x}\right|= \left|\frac{eB_z}{\gamma mv}\right|+O(\epsilon).
\label{rho}
\ee 
Here we proceed to the bunch particles whose position ${\bf r}$ at injection point at $x=0$ is given by
\be
{\bf r}=(\rho+y){\bf e}_2+z{\bf e}_3.
\ee
Thus beam has the initially (y,z) cross section relative to the averaged beam line.

In this set-up, it is more suitable to consider \cite{Silenko},
\bea
&&\bomega=\bomega_s-\bomega_{pz}=-\frac{e}{m}\left[a{\bf B}-\frac{a\gamma }{\gamma+1}({\bf v}\cdot{\bf B}){\bf v}+\left(\frac{1}{\gamma^2-1}-a\right)({\bf v}\times{\bf E})+\frac{1}{\gamma}\left({\bf B}_\parallel-\frac{1}{v^2}({\bf v}\times {\bf E})_\parallel\right)\right.\nonumber\\
&&\left.+\frac{\eta}{2}\left({\bf E}-\frac{\gamma}{\gamma+1}{\bf v}({\bf v}\cdot {\bf E})+({\bf v}\times {\bf B})\right) \right].
\label{Nelsonf2}
\eea
Here $\bomega_{pz}$ is z-component of $\bomega_{p}$ and suffix parallel indicates the projected vector onto the horizontal 
space, that is, ${\bf B}=({\bf B}_{\parallel},~B_z)$.
%In this paper we consider realistic motion in the storage ring where the beam has small width and pitch angles in both (y,z) directions.
So let us consider how Eq.\bref{Nelsonf2} is expressed in these profile \cite{Yokoya}.
\be
{\bf v}=\frac{d{\bf r}}{dt}=\frac{dx}{dt}\frac{d{\bf r}}{dx}=\frac{dx}{dt}\left[\left(1+\frac{y}{\rho}\right){\bf e}_1+y'{\bf e}_2+z'{\bf e}_3\right],
\ee
where $'$ indicates the derivative with respect to $x$. Therefore, the absolute value $v$ of ${\bf v}$ is given by
\be
v=\frac{dx}{dt}\sqrt{\left(1+\frac{y}{\rho}\right)^2+y'^2+z'^2}\equiv \frac{dx}{dt}{\mathscr N}.
\ee
Hereafter, we consider the case in the absence of ${\bf E}$ as in the J-PARC muon g-2/EDM. That is, setting ${\bf E}=0$ in \bref{Nelsonf2}, we obtain
\be
\bomega_J\equiv~(\bomega_s-\bomega_{pz})\mid_{{\bf E}=0}~=-\frac{e}{m}\left[a{\bf B}-\frac{a\gamma }{\gamma+1}({\bf v}\cdot{\bf B}){\bf v}+\frac{1}{\gamma}{\bf B}_\parallel+\frac{\eta}{2}({\bf v}\times {\bf B}) \right].
\label{Nelsonf3}
\ee
%In this case v is conserved. Also we assume $y/\rho,~z/\rho,~y',~z'$ are small quantities of same order $\epsilon$ and herea%fter we take up to $O(\epsilon^2)$. Under these conditions, the particle acceleration is
%\bea
%&&\frac{d{\bf v}}{dt}\approx v^2\left(1-2\frac{y}{\rho}\right)\left[2\frac{y'}{\rho}{\bf e}_1+\left(1+\frac{y}{\rho}\right)\%left(y''-\frac{1}{\rho}\right){\bf e}_2+z''{\bf e}_3\right]\nonumber\\.
%&&\approx v^2\left[2\frac{y'}{\rho}{\bf e}_1+\left(y"-\frac{1}{\rho}+\frac{y}{\rho^2}\right){\bf e}_2+z''{\bf e}_3\right].
%\label{acceleration}
%\eea
%The front factor is corrected to $\left(1-2\frac{y}{\rho}\right)$ from $\left(1-\frac{y}{\rho}\right)$ in \cite{Yokoya} by c%onsidering normalization correctly.
Here
\bea
\label{veq1}
&&{\bf v}\cdot {\bf B}=\frac{v}{{\mathscr N}}\left((1+\frac{y}{\rho})B_x+y'B_y+z'B_z\right),\\
&&{\bf v}\times {\bf B}=\frac{v}{{\mathscr N}}\left[\left(y'B_z-z'B_y\right){\bf e}_1+\left(z'B_x-(1+\frac{y}{\rho})B_z\right){\bf e}_2+\left((1+\frac{y}{\rho})B_y-y'B_x\right){\bf e}_3\right].
\label{veq2}
\eea
Substituting \bref{veq1} and \bref{veq2} into \bref{Nelsonf3}, we obtain
\bea
\label{Nelsonf4}
&&\bomega_J=-\frac{e}{m}\left[\left\{\frac{1}{\gamma}(a+1)B_x{\bf e}_1+\left((a+\frac{1}{\gamma})B_y-\frac{\eta}{2}vB_z\right){\bf e}_2+\left(aB_z+\frac{\eta}{2}vB_y\right){\bf e}_3\right\}\right. \nonumber\\ 
&&-\left\{\left(a(1-\frac{1}{\gamma})(y'B_y+z'B_z)-\frac{\eta}{2}v(y'B_z-z'B_y)\right){\bf e}_1\right.\\
&&\left.\left.+\left(a(1-\frac{1}{\gamma})(B_x+z'B_z)y'+\frac{v\eta}{2}(z'B_x-\frac{y}{\rho})\right){\bf e}_2-a(1-\frac{1}{\gamma})(B_x+z'B_z)z'{\bf e}_3\right\}\right].\nonumber
%\label{Nelsonf4}
\eea
The same type of equation is obtained in \cite{Yokoya} in the absence of EDM. This equation is very important for the experiments measuring anomalous MDM and EDM simultaneously.
\section{Betatron oscillation}
Transversal betatron oscilation equation is derived from the Lorentz equation for the particles with given extent.

Lorentz equation is
\be
\gamma m \dot{\bf v}=e{\bf v}\times {\bf B}.
\label{Lorentzeq}
\ee
Here we assume weak magnetic focusing,
\be
{\bf B}=(0, \frac{\pa B_z}{\pa y}z, B_z+\frac{\pa B_z}{\pa y}y).
\label{weakB}
\ee
We are considering the case ${\bf E}=0$ and $v$ is conserved, that is,
\be
\frac{dv}{dt}=\frac{d^2x}{dt^2}{\mathscr N}+{\mathscr N}'\left(\frac{dx}{dt}\right)^2=0.
\ee
We assume $y/\rho,~z/\rho,~y',~z'$ are small quantities of same order $\epsilon$ and hereafter we take up to $O(\epsilon^2)$. Under these conditions, the particle acceleration is
\bea
&&\frac{d{\bf v}}{dt}\approx v^2\left(1-2\frac{y}{\rho}\right)\left[\frac{y'}{\rho}{\bf e}_1+\left(1+\frac{y}{\rho}\right)\left(y''-\frac{1}{\rho}\right){\bf e}_2+z''{\bf e}_3\right]\nonumber\\.
&&\approx v^2\left[\frac{y'}{\rho}{\bf e}_1+\left(y"-\frac{1}{\rho}+\frac{y}{\rho^2}\right){\bf e}_2+z''{\bf e}_3\right].
\label{acceleration}
\eea
Substiuting \bref{veq2} and \bref{acceleration} into \bref{Lorentzeq}, we obtain
\bea
&&y''=-\frac{1-n(x)}{\rho^2(x)}y, \label{eqmotion1}\\
&&z''=-\frac{n(x)}{\rho (x)^2}z.
\label{eqmotion2}
\eea
Here 
\be
n\equiv -\frac{\rho(x)}{B_z(x,0,0)}\frac{\pa B_z(x,0,0)}{\pa y}\mid_{y=0}=-\frac{e\rho^2}{p}\frac{\pa B_z}{\pa y}.
\ee
It follows from \bref{eqmotion1} and \bref{eqmotion2} that $1>n>0$ must hold for focusing. 
Thus we obtain the transversal betatron oscillation for the given initial extent of bunch beam.
The detailed analyses of the experimental situation are discussed in a separate form.

\section{Discussions}
We have discussed on the spin precession of particles having both anomamalous MDM and EDM in a storage ring.
This is the generalization of \cite{fuku1} to bunch particles having some transversal extent.
Also we have obtained the transversal betatron oscillations for the same objects.
These are very useful for the ongoing and near future experiments measuring anomalous MDM and EDM simultaneously.
For the muon storage ring the observed particles are decayed positrons and we give useful formula for it in appendix A, which gives the relation between the muon spin direction and decayed positron's direction.
So far we have treated particles as classical ones. We show in appendix B that it is suitable.

\section*{Acknowledgements}
We are greatly indebted to Dr. N. Saito and Dr. T. Mibe for useful discussions and hospitality at J-PARC. 
The work of T.F. is supported in part by Grant-in-Aid for Science Research from the Ministry of Education, Science and Culture (No.~26247036). 
\appendix
\include{92_appendix}
\section{Muon decay}
Muon ($\mu^+$) decay ratio, 
\be
\mu^+\rightarrow e^++\nu_e+\overline{\nu}_\mu^,
\ee
is given by \cite{Lifshitz1}
\be
dw=\frac{G_F^2m_\mu^5}{768\pi^4}(1+{\bf \zeta}_e\cdot {\bf n}_e)\left[\left(3-2\frac{\epsilon_e}{\epsilon_{max}}\right)-{\bf\zeta}_\mu\cdot{\bf n}_e\left(1-2\frac{\epsilon_e}{\epsilon_{max}}\right)\right]\frac{\epsilon_e^2d\epsilon_edo_n}{\epsilon_{max}^3}
\ee
with
\be
1+{\bf \zeta}_e\cdot {\bf n}_e=2.
\ee
Here $\epsilon_e$ is the energy of decayed positron and $\epsilon_{max}\approx m_\mu/2$ is the maximal energy of positron. ${\bf n}_e$ is the unit direction vector of positron.
\section{Synchrotron radiation}
Let us consider backgroud in muon storage ring.
Since the distribution function of synchrotron radiation is given by \cite{Lifshitz2}
\be
dI=d\omega \frac{\sqrt{3}}{2\pi}\frac{e^3B}{m_\mu c^2}F\left(\frac{\omega}{\omega_c}\right), ~~F(\xi)\equiv \xi\int_\xi^\infty K_{5/3}(\xi )d\xi,
\ee
where
\be
\omega_c\equiv \frac{3eB}{2m_\mu c}\left(\frac{\epsilon}{m_\mu c^2}\right)^2=\frac{3}{2}\omega_B\gamma^2,
\ee
and $K_\nu$ is the MacDonald function. The value of $F(\xi)$ is shown in Fig.2.
\begin{figure}[h]
\begin{center}
\includegraphics[scale=1.0]{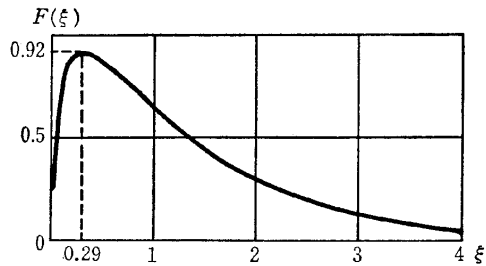}
\end{center}  
%\vspace*{-0.5cm}
\caption{\label{fig:geom_config}
a graph of $F(\xi)$ cited from \cite{Lifshitz3}.}
\end{figure}
Therefore, most of radiation is concentrated around $\omega_c$ and in the plane of the orbit. 
We have treated muons as classical particles moving mainly circular orbit. This is accepted well for the case of J-PARC.
The applicability of classical theory is determined \cite{Lifshitz3}
\be
\chi\equiv \frac{\hbar \omega_B}{\epsilon}\left(\frac{\epsilon}{m_\mu c^2}\right)^3\ll 1.
\ee
This is easily satisfied in J-PARC case. Also 
\be
\hbar \omega_c \ll 2m_ec^2
\ee
and, therefore, there is no creation of positron except for $\mu^+$ decay.

\end{document}